\documentclass[aps,twocolumn,showpacs,superscriptaddress]{revtex4}

\usepackage{graphicx}
\usepackage{latexsym}
\usepackage{amssymb}
\usepackage{amsmath}

\begin{document}

\title{Pressure effect on magnetic susceptibility of SmS in semiconducting phase: experimental study}

\author{A.S. Panfilov}
\affiliation{B. Verkin Institute for Low Temperature Physics and Engineering, 
National Academy of Sciences, 61103 Kharkov, Ukraine}

\author{G.E. Grechnev}
\email[]{grechnev@ilt.kharkov.ua}
\affiliation{B. Verkin Institute for Low Temperature Physics and Engineering, 
National Academy of Sciences, 61103 Kharkov, Ukraine}

\author{D.A. Chareev}
\affiliation{Institute of Experimental Mineralogy, Russian Academy of Sciences, 142432, Chernogolovka,
Moscow District, Russia}
\affiliation{Institute of Physics and Technology, Ural Federal University,
620002 Ekaterinburg, Russia}

\author{A.A. Polkovnikov}
\affiliation{Tyumen' State University, 625003 Tyumen', Russia}

\author{O.V. Andreev}
\affiliation{Tyumen' State University, 625003 Tyumen', Russia}


\begin{abstract}
Magnetic susceptibility $\chi$ of the polycrystalline sample of samarium monosulfide 
was measured  as a function of the hydrostatic pressure $P$ up to 2 kbar at 
liquid nitrogen and room temperatures using a pendulum-type magnetometer. 
A pronounced magnitude of the pressure effect is found to be
positive in sign and strongly temperature dependent: 
the pressure derivatives of $\chi$, d\,ln$\chi$/d$P$, are 
$6.3\pm0.5$ and $14.2\pm1$ Mbar$^{-1}$ at 300 and 78 K, respectively.
The obtained experimental results are discussed within phenomenological approaches.
\end{abstract}


\maketitle

\section{Introduction}

SmS is a typical representative of the samarium compounds with an unstable 4f shell.
At ambient pressure SmS is a black-colored narrow-gap semiconductor with nearly 
divalent Sm$^{2+}$ ions in 4f$^6$($^7F_0$) ground state configuration. 
At a critical pressure $P_c\sim 6.5$~kbar and room temperature SmS undergoes 
the first order isostructural phase transition to a mixed-valent metal state 
accompanied by a significant volume collapse of about 13~\% and a color change 
from black to golden-yellow \cite{Jayaraman70b,Jayaraman74,Maple71,Smirnov78}. 
According to theoretical calculations of the volume dependent electronic structure 
of SmS, SmSe an SmTe (see, for example, \cite{Svane04,Svane05,Gupta09,Li14}) 
this transition, being typical for the whole series 
\cite{Jayaraman70b,Smirnov78,Jayaraman70a,Sidorov89}, is related 
to the closure under pressure of the semiconducting gap $\Delta$, 
whose values at zero pressure are listed in table \ref{Summary}.
\begin{table}[h]
\caption{\label{Summary} Summary of some relevant properties of Sm chalcogenides: 
semiconducting energy gap $\Delta$ \cite{Wachter94}, 
pressure derivative d$\Delta$/d$P$ \cite{Jayaraman74}, 
lattice parameter $a$ at room temperature \cite{Bucher71}, 
experimental values of the magnetic susceptibility $\chi(0)$ at $T\to 0$ K 
with the subtracted impurity effects \cite{Bucher71,Birgeneau72}.}
\vspace{3pt}
\begin{center}
\begin{tabular}{c|ccc}
\hline\hline
Property &~~~~~~~SmS~~~~~&~~~SmSe~~~&~~~SmTe~~~\\
\hline
$\Delta$, eV & 0.15, $0.2^a$      &  0.45    &   0.65   \\
             & $0.23^b$, $0.18^c$ & $0.47^c$ & $0.67^c$ \\
d$\Delta$/d$P$, eV/kbar   & $-0.010$~~ & $-0.011$~~ & $-0.012$~~ \\
$a$, \AA  & 5.970 &  6.202 & 6.601\\
~~$\chi(0)$, $10^{-3}$ emu/mol~~         & 9.15 &  7.85 & 7.0 \\
\hline \hline
\multicolumn{4}{l}{$^a$Ref. \cite{Bucher71},~~~$^b$Ref. \cite{Kaminskii12},~~~$^c$
calculated value from Ref. \cite{Antonov02}}
\end{tabular}
\end{center}
\end{table}

Because the semiconductor-metal transition in SmS is the most pronounced 
and its critical pressure $P_c$ is relatively small, there is an opportunity 
to study this transition in details by measuring the magnetic properties, 
which are very sensitive to the Sm valence state.
The first and, in fact, the only study of the pressure effect on magnetic susceptibility $\chi$
of SmS was performed at room temperature and pressure $P$ up to 18 kbar by Maple and Wohlleben
\cite{Maple71}. 
They observed a distinct drop in $\chi(P)$ at $P_c$ by about 65\% 
within 0.5 kbar and further decrease of $\chi$ with increasing $P$, 
but at a much slower and decreasing rate. 
In the context of the present report, we would like to note observation 
of a marked increase in susceptibility of semiconducting SmS at low pressures, which is 
described by the derivative d$\chi$/d$P=+(78.3\pm7)\times 10^{-6}$ emu/mol$\cdot$kbar. 
An initial increase in $\chi$ at room temperature with d$\chi$/d$P=17\times10^{-6}$ and 
$2.2\times10^{-6}$ emu/mol$\cdot$kbar was also reported for SmSe and SmTe, respectively \cite{Maple71}.

As it follows from comparison of the experimental values of magnetic susceptibility 
at $T=0$ K for SmS, SmSe and SmTe with the Van Vleck susceptibility of Sm$^{2+}$ 
free ions at zero temperature, $\chi_{\rm VV}(0)\sim7.1\times10^{-3}$ emu/mol \cite{Birgeneau72}, 
the Van Vleck contribution in $\chi$ of SmX is the basic one. 
Its value is determined as
\begin{equation}
\chi_{\rm VV}(0)=8N\mu_{\rm B}^2/E_{so},
\label{X_VV}
\end{equation}
where $E_{so}$ is the multiplet splitting, $E(^7F_0)-E(^7F_1)\simeq422$~K \cite{Dupont67}.
Since this contribution has the intra-atomic origin and would weakly depend 
on interatomic distance, the observed large pressure effect on $\chi$ in SmS 
and related compounds is, at first glance, unexpected and surprising. 
It should be noted that the magnitude of this effect correlates reasonably 
with the value of an excess susceptibility, $\delta\chi=\chi(0)-\chi_{\rm VV}(0)$.
Besides, as can be seen, the value of $\chi$ (and $\delta\chi$) in the series  
SmTe$\to$SmSe$\to$SmS increases with reduction of $\Delta$, and the positive sign 
of the pressure effect on $\chi$ coincides with experimentally observed decrease 
of $\Delta$ under pressure \cite{Jayaraman74}. 
Based on the foregoing, one can assume that the excess susceptibility term 
$\delta\chi$, being very sensitive to applied pressure, is closely related 
to the electronic structure and its changes under pressure.
Because of the high sensitivity to interatomic distances, the results of 
the experimental studies of magnetic properties under pressure provide 
the efficient tool to verify the theoretical descriptions of magnetism 
and its relation to electronic structure of samarium chalcogenides.

In this paper we have carried out the precise study of the pressure effect 
on magnetic susceptibility of semiconducting SmS in order to complete and 
validate  the previous experimental data \cite{Maple71}. 
The obtained results have been compared with the predictions of the existing model approaches.

\section{Experimental details and results}

Polycrystalline samples of samarium monosulfide were prepared by 
the method described in Ref.~\cite{Andreev93}. 
A powder x-ray diffraction analysis revealed that the samples 
consist mainly of the NaCl phase of SmS ($\sim 95\%$) and some amount 
of Sm$_3$S$_4$ and Sm$_2$O$_2$S impurity phases.

\begin{figure}[]
\begin{center}
\includegraphics[width=0.4 \textwidth]{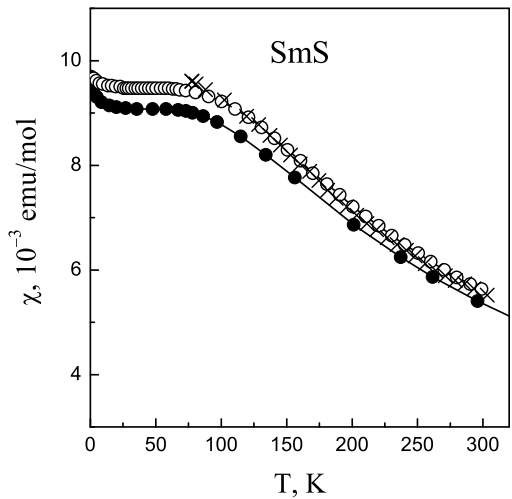}
\caption{Temperature dependence of magnetic susceptibility of SmS: 
{\large$\circ$} -- data for the single crystalline sample from \cite{Matsubayashi07}, 
{\large$\bullet$} -- \cite{Bucher71}, $\times$ -- our data.} \label{X(T)}
\end{center}
\end{figure}

The temperature dependence of magnetic susceptibility $\chi(T)$ of SmS was measured 
by a Faraday method in the temperature range of $78-300$~K at magnetic field of $H=5$~kOe. 
As is seen in fig.~\ref{X(T)}, the observed dependence is in close agreement 
with literature data for SmS that testifies to rather high quality of the prepared 
sample and a negligible contribution of impurity phases to its magnetism.

The precise study of the uniform pressure effect on magnetic susceptibility of SmS was carried
out under helium gas pressure $P$ up to 2 kbar, using a pendulum type magnetometer.
The measured sample was placed inside a small compensating coil located at the
lower end of the pendulum rod. 
Then under switching on magnetic field, the value of current through the coil, 
at which the magnetometer comes back to its initial position, is the measure 
of the sample magnetic moment. 
To measure the pressure effects, the pendulum magnetometer was inserted into 
a  cylindrical nonmagnetic pressure chamber which was in turn placed into a cryostat. 
Measurements were performed at fixed temperatures, 78 and 300 K, to eliminate the effect 
on susceptibility of the temperature change during applying or reversing pressure. 
For detailed description of the device and analysis of the sources of 
the experimental errors, see Ref.~\cite{Panfilov15}.
In our case, the relative errors of measurements of $\chi$ under pressure did not exceed
0.1\% for the employed magnetic field $H=17$~kOe.

\begin{figure}[t]
\begin{center}
\includegraphics[width=0.4 \textwidth]{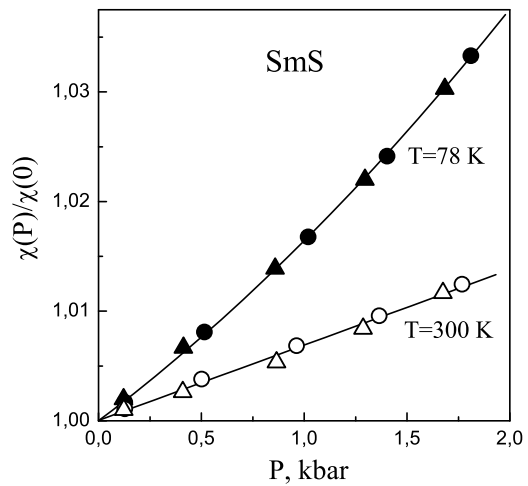}
\caption{Pressure dependence of magnetic susceptibility for SmS at 78 and 300 K. 
The circ and triangle symbols correspond to increasing and decreasing pressure, 
respectively.} \label{X(P)}
\end{center}
\end{figure}

The experimental $\chi(P)$ dependencies,
normalized to value of $\chi$ at zero pressure, are presented in fig.~\ref{X(P)}.
To check the data reliability, the $\chi(P)$ dependencies were measured for two samples 
with different mass (120 and 55 mg) and appeared to be identical within experimental errors.

As seen in fig.~\ref{X(P)}, the $\chi(P)$ behavior shows a pronounced increase 
in $\chi$ under pressure. 
It should be noted that a weak nonlinearity with positive curvature of 
$\chi(P)$ dependence is observed for $T=78$~K. 
For $T=300$~K the pressure effect is linear within experimental errors. 
The corresponding values of the pressure derivative 
d\,ln$\chi$/d$P$ $(\equiv (\Delta\chi/\chi)/\Delta P,~{\rm at}~ \Delta P\to 0 )$ 
are listed in table~\ref{exp} together with the values of $\chi$ at $P=0$.

\begin{table}[h]
\caption{Magnetic susceptibility $\chi$ and its pressure derivative 
d\,ln$\chi$/d$P$ (at $P\to 0$) at different temperatures.}
\vspace{5pt} \label{exp}
\begin{center}
\begin{tabular}{ccc}
\hline
 T (K)&~~ $\chi$ ($10^{-3}$ emu/mol)~~ &~~ d\,ln$\chi$/d$P$ (Mbar$^{-1}$)\\
\hline
  78 & 9.51  & $ 14.2 \pm 1$ \\
\vspace {5 pt}
 300 & 5.56  & $ 6.3\pm0.5$ \\
\hline
\end{tabular}
\end{center}
\end{table}

\begin{figure}[]
\begin{center}
\includegraphics[width=0.4\textwidth]{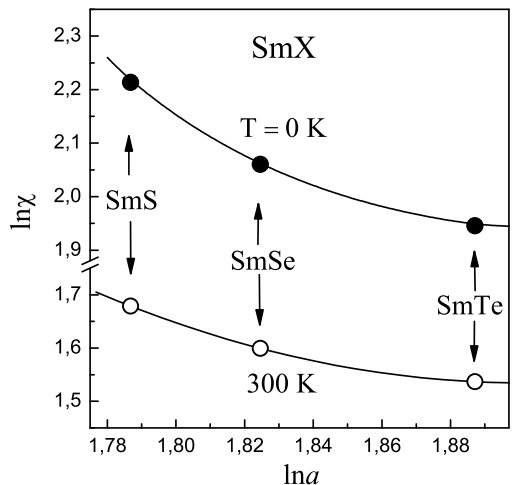}
\caption{Magnetic susceptibility of SmX (X=S, Se and Te) versus lattice parameter 
$a$ in logarithmic presentation, using the experimental data collected in Ref.~\cite{Bucher71}.}
\label{X(a)}
\end{center}
\end{figure}

\section{Discussion}

Firstly we note without comments a quantitative difference between the pressure 
effect value at room temperature obtained here, d\,ln$\chi$/d$P\simeq6.3$ Mbar$^{-1}$, 
and that resulted from the experimental data of Ref.~\cite{Maple71}, 
d\,ln$\chi$/d$P\simeq 15$ Mbar$^{-1}$.
In either case, the initial increase of the magnetic susceptibility with pressure 
correlates qualitatively with behavior of the zero-pressure susceptibility of SmS, 
SmSe and SmTe compounds as a function of their lattice parameter $a$ 
at ambient conditions (fig.~\ref{X(a)}). 
By this way, the effect of chemical pressure can be estimated as
\begin{equation}
{\partial{\rm \,ln}\chi\over\partial P}\sim 4.0~{\rm Mbar}^{-1}~~~{\rm and} \sim1.8~  {\rm Mbar}^{-1}
\label{dX/da}
\end{equation}
at $T=0$ and 300~K, respectively. using the corresponding values of
$\partial$\,ln$\chi$/$\partial$\,ln$V\sim-2.0$ and $\sim-0.9$, 
which follow from fig.~\ref{X(a)}, and the bulk modulus value for 
SmS, $B\simeq 0.5$ Mbar \cite{Hailing84}. 
As is seen from (\ref{dX/da}) and table~\ref{exp}, the estimated value of the 
chemical pressure effect is about 30\% of the external pressure effect both at low 
and at room temperatures, giving evidence for its strong temperature dependence. 
Besides, it is obvious that difference in the magnetic properties along
the SmX family is governed not only by variations in the interatomic distance, 
but also the chemical nature of the anion.

There are a few approaches to explain the positive pressure effect on magnetic
susceptibility for semiconducting SmS.
For example, Maple and Wohlleben \cite{Maple71} assumed that this effect 
arises from a decrease under pressure of the multiplet splitting 
$E_{so} =E(^7F_0)-E(^7F_1)$ of the Sm$^{2+}$ ion due to increased screening 
of the nuclear charge, caused by increasing hybridization of the 4f states 
with d states of the Sm ion across the decreasing semiconducting energy gap $\Delta$. 
However, this assumption is hardly consistent with a commonly believed proximity of the 
$E_{so}$ energies along the series of SmS, SmSe and SmTe compounds 
\cite{Zelezny89,Nathan75,Shapiro75} which have significantly different values 
of the lattice parameter and semiconducting gap $\Delta$. 
In addition, as it follows from neutron-scattering studies in the semiconducting 
phase of SmS \cite{McWhan78}, there is no observable change in the value of 
$E_{so}$ under pressure, which is  close to that for the free Sm$^{2+}$ ion.
Therefore, it seems reasonable to conclude that the Van Vleck contribution 
of Sm ions is pressure independent and presumably the same in magnitude for the SmX series.

To explain the origin of the observed excess term in susceptibility of Sm chalcogenides, 
Birgeneau {\em et~al.} \cite{Birgeneau72} ascribed it to effect of the Sm--Sm exchange 
interaction, which may be treated in the molecular field approximation. 
Then susceptibility at $T=0$~K is given by
\begin{equation}
\chi(0)=8N\mu_{\rm B}^2/(E_{so}-\Theta)\approx\chi_{\rm VV}(0)[1+\Theta/E_{so}].
\label{exchange}
\end{equation}
Here $\Theta$ is the effective exchange parameter and this approximation is valid at $\Theta/E_{so}<<1$.
From analysis of the susceptibility data, authors of Ref. \cite{Birgeneau72} 
have estimated the exchange parameter and its variation with chemical pressure. 
It was concluded \cite{Birgeneau72} that the obtained information provides 
a remarkable description of the pressure dependence of the SmX susceptibility 
\cite{Maple71} within the proposed simple theory.
However, it should be noted that i) the chemical pressure effect differs 
essentially from that of the external pressure, ii) the estimated values of 
d$\chi$/d$P$ at $T=0$ K \cite{Birgeneau72} were unreasonably compared 
with experimental data \cite{Maple71} at room temperature. 
Besides, an origin of the strong dependence of exchange interaction 
on the interatomic distance is not clear.

\begin{figure}[h]
\begin{center}
\includegraphics[width=0.4\textwidth]{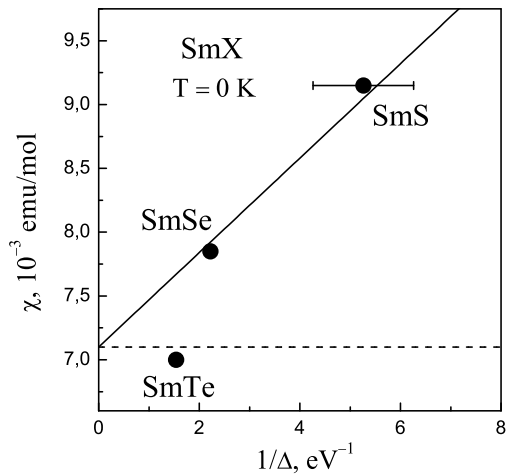}
\caption{Magnetic susceptibility of SmS, SmSe and SmTe at $T\to 0$~K \cite{Bucher71} 
versus the inverse semiconducting gap $\Delta$. 
The dashed line marks the Van Vleck susceptibility of the free Sm$^{2+}$ ion 
at $T=0$~K, the solid line is a guide for the eye.}
\label{X(Eg)}
\end{center}
\end{figure}

Below we propose an alternative description of magnetic properties of 
the SmX series, based on the data in figure \ref{X(Eg)}, where the 
magnetic susceptibility at zero temperature is presented as a function 
of the inverse value of semiconducting gap of the compounds. 
As is seen, the susceptibility can be satisfactorily described, excepting the data for SmTe, as
\begin{equation}
\chi(0)\simeq \chi_{\rm VV}(0)+A/\Delta
\label{X_band}
\end{equation}
with $A\simeq0.37\times10^{-3}$ (emu/mol)$\cdot$eV. 
Then resulted from Eq. \ref{X_band} value of the pressure derivative of $\chi(0)$ is given by
\begin{equation}
{{\rm d}\chi(0)\over{\rm d}P}\simeq -{A\over\Delta^2}\times{{\rm d}\Delta\over{\rm d}P},
\label{dX/dP}
\end{equation}
where pressure dependencies of $\chi_{\rm VV}(0)$ and parameter $A$ are neglected. 
Substitution in Eq. \ref{dX/dP} the mean value $\Delta=0.19$ eV and 
d$\Delta$/d$P=-0.010$ eV/kbar from table \ref{Summary} yields for SmS a value 
of the pressure derivative, d$\chi(0)$/d$P\simeq 102\times10^{-6}$ emu/mol$\cdot$kbar. 
The analogous estimate for SmSe equals to d$\chi(0)$/d$P\simeq 19\times10^{-6}$ emu/mol$\cdot$kbar. 
The obtained estimates of d$\chi(0)$/d$P$ at $T=0$ K agree reasonably with our 
experimental value for SmS at $T=78$ K, d$\chi$/d$P\simeq135\times10^{-6}$ emu/mol$\cdot$kbar, 
and data for SmSe at room temperature from Ref. \cite{Maple71}, 
d$\chi$/d$P\simeq17\times10^{-6}$ emu/mol$\cdot$kbar. 
This fact indicates the dominating contribution of the term $A/\Delta$ in 
Eq. \ref{X_band} to the behavior of the susceptibility under pressure.
Based on a type of the functional dependence, we suggest the excess susceptibility 
$A/\Delta$ to be the orbital magnetism of the band electrons of the Van Vleck type.

\section{Conclusion}

Our precise measurements of the magnetic susceptibility under pressure for semiconducting SmS
have revealed the pronounced positive pressure effect, being strongly temperature dependent.

Analysis of the obtained experimental results within approach, based on 
a phenomenological Heisenberg exchange between Sm$^{2+}$ ions \cite{Birgeneau72}, 
gives, in our opinion, only qualitative agreement with experiment. 
Besides, it requires a plausible explanation of the strong dependence 
of exchange interaction on the interatomic distance.

Presented here the alternative model assumes two main contributions 
to the magnetic susceptibily of Sm chalcogenides -- the pressure-independent 
Van Vleck susceptibility of Sm$^{2+}$ ions and the excess susceptibility, 
which is inversely proportional to the value of semiconducting gap $\Delta$ 
and strongly dependent on pressure.
Within this approach we have obtained a reasonable description of the experimental 
data using the only experimental value of the pressure derivative d$\Delta$/d$P$.

In addition, the excess susceptibility term is tentatively suggested to arise 
from the orbital magnetism of the band electron states. 
To verify this suggestion the theoretical calculation of this contribution have to be carried out.

\section*{Acknowledgement}
The work was partly supported by Act 211 Government of the Russian Federation, agreement N 02.A03.21.0006

\end{document}